\documentclass[prl,twocolumn,showpacs,superscriptaddress,amsmath,amssymb,reprint]{revtex4}

\usepackage{graphicx}
\usepackage{latexsym}
\usepackage{amsmath}
\usepackage{amssymb}
\usepackage{amsfonts}
\usepackage{color}

\newcommand{\ds}{\displaystyle}

\begin{document}
\bibliographystyle{apsrev}

\title{Spontaneous currents in superconducting systems with strong spin-orbit coupling}

\author{S. Mironov}
\affiliation{Institute for Physics of Microstructures, Russian Academy of Sciences, 603950 Nizhny Novgorod, GSP-105, Russia}
\author{A. Buzdin}
\affiliation{University Bordeaux, LOMA UMR-CNRS 5798, F-33405 Talence Cedex, France}

\date{\today}
\begin{abstract}
We show that Rashba spin-orbit coupling at the interface between a superconductor and a ferromagnet should produce a spontaneous current in the atomic thickness region near the interface. This current is counter-balanced by the superconducting screening current flowing in the region of the width of the London penetration depth near the interface. Such current carrying state creates a magnetic field near the superconductor surface, generates a stray magnetic field outside the sample edges, changes the slope of the temperature dependence of the critical field $H_{c3}$ and may generate the spontaneous Abrikosov vortices near the interface.
\end{abstract}


\maketitle

The influence of the strong spin-orbit coupling (SOC) on superconducting systems stays in the focus of intensive research for more than two decades (see as reviews \cite{Mineev_Review} and \cite{Agterberg_review}). In the absence of the inversion symmetry an electron spin $\vec{\sigma}$ becomes coupled with the orientation of the momentum $\vec{p}$, which produces the non-trivial ``helicity'' of the electronic energy bands. The resulting helical states \cite{Edelstein_HelicalStates, Mineev_HelicalStates} play the central role in the appearance of Majorana modes \cite{Alicea}, the formation of Josephson $\varphi_0$-junctions with spontaneous phase difference in the ground state \cite{Buzdin_Phi, Krive, Reynoso, Kouwenhoven}, and the emerging of different types of superconducting phases with finite Cooper-pair momentum which are similar to the Fulde-Ferrell-Larkin-Ovchinnikov (FFLO) states \cite{Agterberg_review}.

\begin{figure*}[t]
\includegraphics[width=0.8\textwidth]{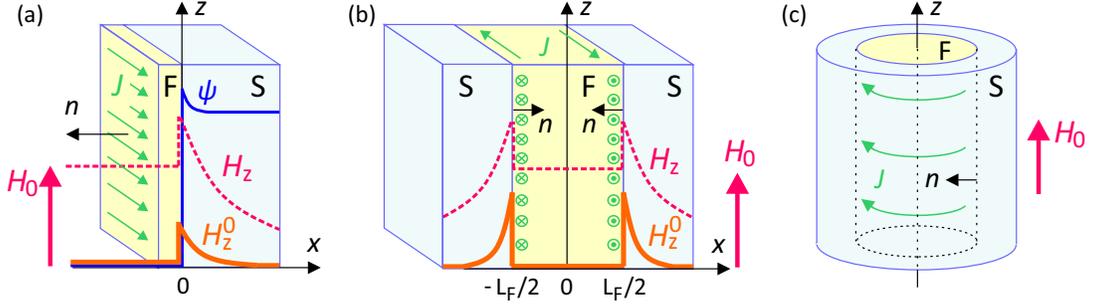}
\caption{(Color online) Hybrid superconductor (S) - ferromagnet (F) systems where the spin-orbit coupling produces spontaneous currents: (a) S/F bilayer, (b) S/F/S sandwich, and (c) bulk superconductor with the cylindrical ferromagnetic core. The direction of the spontaneous current ${\bf J}$ is shown with the green arrows, the corresponding profile of the spontaneous magnetic field in the absence of external magnetic field is plotted schematically with the orange color, the distribution of the order parameter $\psi$ is shown with the blue curve. The magenta curves show the magnetic field profiles in the presence of the external magnetic field ${\bf H}_0$ directed along the $z$-axis. The unit vector {\bf n} is the vector in the direction of the broken inversion symmetry at the S/F interfaces which determines the energy of the spin-orbit coupling.} \label{Fig_System}
\end{figure*}

One of the key questions related to the physics of superconducting systems with broken inversion symmetry is the existence of the spontaneous electric current originating from the interplay between SOC and magnetic order. Indeed, the SOC of the Rashba type in the non-centrosymmetric metals provides the additional contribution $\propto(\vec{\sigma}\times \vec{p})\cdot \vec{n}$ to the electron energy ($\vec{n}$ is the unit vector along the axis with the broken inversion symmetry). The ferromagnetic order or strong external magnetic field polarizes the electron spins making the momentum direction along the vector $\vec{\sigma}\times\vec{n}$ energetically more favorable compared with others, which suggests the possibility to have a spontaneous electric current.

The detailed analysis shows that the situation is more subtle and typically no spontaneous current appear. In bulk materials the appearance of the current-carrying states are unfavorable because of the large corresponding kinetic energy of the condensate. For two-dimensional superconductors the SOC induces several types of FFLO-like helical phases with non-zero Cooper-pair momentum $\vec{p}$ in the ground state \cite{Barzykin, Samokhin_SOC, Kaur_SOC, Edelstein_PRL, Dimitrova}. It was claimed that in the presence of the in-plane magnetic field such states can carry the supercurrent \cite{Edelstein_Current}. However, the accurate analysis shows that in all mentioned situations the Cooper pair wave function $\psi\propto e^{i\vec{p}\vec{r}}$ does not produce the electric current since the SOC modifies the quantum-mechanical expression for the current by adding the terms which exactly compensate the usual orbital contribution \cite{Mineev_HelicalStates, Agterberg}. Note that it has been predicted that in the unconventional d-wave and chiral p-wave superconductors or at the interfaces between the s-wave superconductors and half-metals the appearance of the Andreev edge states may lead to the ground state with broken time-reversal symmetry \cite{Fogelstrom1997, Vorontsov2009, Hakansson2015, Barash2000, Higashitani, Honerkamp, Fauchere, Bobkova, Matsumoto, Stone, Kwon}. The transition to these states typically occurs well below $T_c$ and may be accompanied by the spontaneous current generation.

In this paper we demonstrate that local Rashba SOC produces the spontaneous currents flowing along the surface of the bulk s-wave superconductors provided this surface is put in contact with a layer of a ferromagnetic insulator. In contrast with the surface magnetism in unconventional superconductors, in our case the spontaneous currents appear at the superconducting transition. Remarkably, the emergence of such current does not require the presence of the external magnetic field and is controlled by the exchange field inside the ferromagnet and the strength of the SOC. Experimentally the spontaneous current is shown to reveal through the appearance of the magnetic field near the interface which can be detected in the local probe measurements and also changes the behavior of the critical field $H_{c3}$ with the variations of temperature $T$. Specifically, the slope of the dependence $H_{c3}(T)$ becomes dependent on the relative orientation between the external magnetic field and the exchange field in the ferromagnet. Also the spontaneous current can serve as a probe of SOC and we may expect that the appropriate conditions to observe this effect should be realized at the interface of ferromagnetic insulator and a superconductor with large nuclear charge Z, like Pb or Hg. Note that recently the unusual enhancement of superconductivity by a parallel magnetic field was observed in thin Pb film \cite{Gardner}. Following the authors, the most probable mechanism of this phenomenon is related with a large SOC in Pb.

To describe the physics of these phenomena use the Ginzburg-Landau (GL) model which is relevant at temperatures $T$ close to the superconducting transition temperature $T_c$. In the presence of Rashba SOC the density $f({\bf r})$ of the GL free energy $F=\int f({\bf r})d^3{\bf r}$ reads \cite{Samokhin_SOC, Kaur_SOC} (we use the system of units where $\hbar=c=1$)
\begin{equation}\label{GL_FE_general}
\begin{array}{c}{\ds
f({\bf r})=a|\psi|^2+\gamma|\hat{\bf D}\psi|^2+\frac{b}{2}|\psi|^4+\frac{({\rm rot}{\bf A})^2}{8\pi}}\\{}\\{ \ds ~~~~~~~~~~ + (\vec{n}\times \vec{h})\cdot[\psi^*\varepsilon({\bf r})\hat{\bf D}\psi+{\rm h.c.}].}
\end{array}
\end{equation}
Here $a=-\alpha(T_c-T)$, $b$ and $\gamma>0$ are the standard GL coefficients, $\psi=|\psi|e^{i\varphi}$ is the superconducting order parameter with $|\psi|^2=|a|/b$, $\hat{\bf D}=-i\nabla+2e{\bf A}$ is the gauge-invariant momentum operator (here $e>0$), $\vec{n}$ is the unit vector in the direction along which the inversion symmetry is broken, $\vec{h}$ is the exchange field, and $\varepsilon({\bf r})$ is the Rashba SOC constant which is nonzero only inside the narrow region near the sample surface. We assume that $h$ strongly exceeds the Zeeman splitting energy due to the external field so that the renormalization of $h$ in (\ref{GL_FE_general}) due to magnetic field can be neglected.

The appearance of the spontaneous magnetic field is a generic phenomenon revealing for a wide class of S/F interfaces with SOC in different superconducting hybrids (see Fig.~\ref{Fig_System}). We start from the simplest situation when a ferromagnetic (F) film is deposited on the surface of the half-infinite superconductor occupying the region $x>0$ (Fig.~\ref{Fig_System}a) so that the inversion symmetry is broken in the $x$-direction and $\vec{n}=-\hat{\bf x}$. When the exchange field in the F-layer has only the in-plane component $\vec{h}=h_z\hat{\bf z}$ the vector product $(\vec{n}\times \vec{h})=h_z\hat{\bf y}$ is also parallel to the superconductor surface. We choose the external magnetic field ${\bf H}_0$ to be directed along the $z$-axis: ${\bf H}_0=H_{0z}\hat{\bf z}$. For simplicity we do not account the inverse proximity effect neglecting the spatial variations of $|\psi|$ in the S-layer and also choose the gauge of the vector potential ${\bf A}$ in a way that $\nabla\varphi=0$. This is justified in the case of the ferromagnetic insulator or more generally when the conductivity of the ferromagnet is much smaller than the normal state conductivity of the superconductor \cite{Kupriyanov}.

Assuming that the SOC is generated only inside the layer of the thickness $l_{so}$ (we may expect that it is nm scale) which is much smaller than the coherence length $\xi=\sqrt{\gamma/|a|}$ we rewrite the last term in Eq.~(\ref{GL_FE_general}) as a surface contribution to the free energy
\begin{equation}\label{F_SO_general}
F_{SO}= 2|\psi|^2\varepsilon l_{so} S~ (\vec{n}\times \vec{h}) \cdot\left.\left(\nabla \varphi+2e {\bf A}\right)\right|_{x=0},
\end{equation}
where $S$ is the area of the sample surface. The derivative of $F_{SO}$ over ${\bf A}$ defines the surface supercurrent ${\bf J}$ originating from the SOC:
\begin{equation}\label{Current_general}
{\bf J}=-\frac{1}{S}\frac{\delta F_{SO}}{\delta {\bf A}}= - \frac{1}{4\pi \lambda^2}\alpha_{so}h_z\hat{\bf y},
\end{equation}
where we introduce the SOC constant $\alpha_{so}=\varepsilon l_{so}/(2 e \gamma)$ and the London penetration depth $\lambda=(32\pi e^2\gamma|\psi|^2)^{-1/2}$. Remarkably, the emergence of the spontaneous current ${\bf J}$ does not require the presence of the external magnetic field, it is a direct consequence of the interplay between the exchange field and the Rashba SOC.

The crucial difference between this result and the situation described in Ref.~\cite{Agterberg} is that in our case the field ${\vec h}$ has the exchange nature and does not depend on the vector potential. In contrast, when the Zeeman splitting of the energy bands for the spin-up and -down electrons is caused by the magnetic field ${\bf H}={\rm rot}{\bf A}$ the expression for the surface current analogous to (\ref{Current_general}) contains an additional term coming from the derivative of ${\bf H}$ over ${\bf A}$ which exactly compensates the contribution (\ref{Current_general}) (see Ref.~\cite{Agterberg}).

According to the Maxwell equations the surface current ${\bf J}$ produces the magnetic field ${\bf H}$ which decays at the scale $\propto \sqrt{S}$. However outside the superconductor this field should be compensated by the field induced by the screening Meissner current so that
\begin{equation}\label{H_profile}
H_z(x)=\left\{\begin{array}{l}{H_{0z}~~~~~~~~~~~~~~~~~~~~~~~~~~~~{\rm for}~x<0,}\\{ (H_{0z}+\Delta H){\rm exp}(-x/\lambda)~~~{\rm for}~x>0,}\end{array}\right.
\end{equation}
where $\Delta H=\alpha_{so}h_z/\lambda^2$ is the magnetic field step due to the surface current ${\bf J}$.

Note that the finite thickness $L_S\gg l_{so}$ of the S film does not substantially change the described phenomenon. In particular, if both surfaces of the superconductor are covered by the identical F-layers with strong SOC the magnetic field inside the S film (for $|x|<L_S/2$) reads $H_z(x)=(H_{0z}+\Delta H)\cosh(x/\lambda)\cosh^{-1}(L_S/2\lambda)$, where the maximal value of $H_z$ is determined by the SOC.

The spontaneous current gives rise to a slight variation of the order parameter near the sample surface. Indeed, taking the vector potential in the S-layer in the form $A_y(x)=-\lambda\Delta H{\rm exp}(-x/\lambda)$ for $H_{0z}=0$ one can check that the surface free energy due to SOC is $F_{SO}=-\lambda S\Delta H^2 /4\pi$. This energy gain is proportional to $|\psi|^3$ which makes the local increase of the order parameter favorable. To calculate $\psi(x)$ we assume that $\lambda \gg \xi$ and in the region where $|\psi|$ deviates from the equilibrium value $|a|/b$ the spatial variation of the magnetic field is negligibly small. The calculation shows \cite{SuppMat}
\begin{equation}\label{Psi_res}
\psi(x)=\psi_0\left(1+\frac{3}{4}\frac{\lambda}{\xi}\frac{\Delta H^2}{H_{cm}^2}e^{-\sqrt{2}x/\xi}\right),
\end{equation}
where $H_{cm}=(2\sqrt{2}e\xi\lambda)^{-1}$ is the thermodynamic critical field. Clearly, the obtained increase of the order parameter arises only at temperatures well below $T_c$ while at the transition temperature the effect vanishes. Note that the most favorable conditions for the growth of $\psi$ are realized in the case of negligibly small inverse proximity effect.

\begin{figure}[t!]
\includegraphics[width=0.3\textwidth]{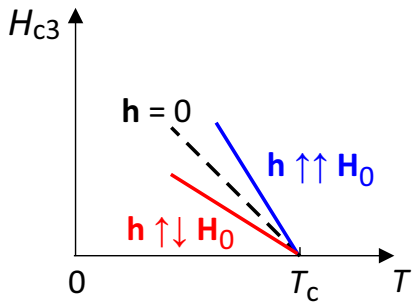}
\caption{(Color online) The behavior of the critical field $H_{c3}$ in the presence of the ferromagnetic layer with the spin-orbit coupling in the case $\varepsilon>0$. The red and blue curves correspond to the different relative orientation between the exchange field ${\bf h}$ and the external field ${\bf H}_0$ while the black dashed line shows the field $H_{c3}$ in the absence of the F layer. } \label{Fig_Hc3}
\end{figure}

In type-II superconductors the spontaneous surface currents substantially change the behavior of the critical magnetic field $H_{c3}$ which corresponds to the emergence of the localized superconducting states above the bulk upper critical field. The SOC and the exchange field in the F-layer produce an additional current which interferes with the usual orbital one screening the external magnetic field and affects the conditions for the superconductivity nucleation. The detailed calculations in the spirit of Ref.~\cite{SG_book} show that the dependence $H_{c3}(T)$ changes its slope near $T_c$ and for $H_{0z}>0$ we find \cite{SuppMat}
\begin{equation}\label{Hc3_res}
H_{c3}(T)\approx H_{c3}^0(T)\left(1+\zeta\varepsilon h_z l_{so}/\gamma\right),
\end{equation}
where $H_{c3}^0(T)=1.6946 (\alpha/2e\gamma)\left(T_c-T\right)$ is the standard dependence of the critical field in the absence of the SOC and $\zeta=1.9847$. Remarkably, the sign of the deviation from $H_{c3}^0(T)$ is determined by the relative orientation of the magnetic field ${\bf H}_0$ and the exchange field ${\bf h}$ (see Fig.~2). Thus, despite the effect of the SOC is typically small it can be observed experimentally by inverting the direction of the magnetic field.

Even more interesting situation occurs when the F-layer of the thickness $L_F$ is placed between two bulk superconductors (see Fig. 1b). Assuming that the SOC is non-zero only at the region of the thickness $l_{so}$ near each S/F interface and $L_F\gg l_{so}$ one finds that inside the superconductors (for $|x|>L_F/2$) the magnetic field has the form $H_z(x)=(H_F+\Delta H){\rm exp}\left[-(|x|-L_F/2)/\lambda\right]$ while inside the F layer (for $|x|<L_F/2$) the field intensity $H_z(x)=H_F$. Here the constant $H_F$ should be defined from the minimization of the Gibbs free energy
\begin{equation}\label{SFS_FE}
G=\int\limits_{-\infty}^\infty \left[H_z^2+\lambda^2\left(\partial_x H_z\right)^2-2H_zH_{0z}\right]\frac{S dx}{8\pi}+F_{SO},
\end{equation}
where $F_{SO}=-2\lambda S\Delta H(H_F+\Delta H)/4\pi$ (here we account that the spontaneous current appears at both S/F interfaces). Note that the magnetization ${\bf M}$ inside the F-layer makes the difference between the magnetic intensity ${\bf H}$ which enters Eq.~(\ref{SFS_FE}) and the magnetic induction ${\bf B}={\bf H}+4\pi{\bf M}$ which is non-zero even if $H_F=0$. Performing the integration we find:
\begin{equation}\label{SFS_FE_res}
G(H_F)=G_1+S\left(H_F^2-2H_FH_{0z}\right)\left(L_F+2\lambda\right)/8\pi,
\end{equation}
where $G_1=-(S\Delta H\lambda/4\pi)\left(\Delta H+2H_{0z}\right)$ does not depend on $H_F$. Interestingly, the minimum $G_{min}$ of the function (\ref{SFS_FE_res}) corresponds to $H_F=H_{0z}$ and $G_{min}=-(SL_F/8\pi)\left[H_{0z}^2+(2\lambda/L_F)(H_{0z}+\Delta H)^2\right]$. This value is negative for any arbitrary $H_{0z}$ and, thus, the state with the spontaneous current is always favorable at $T<T_c$.

Finally, we consider the peculiar situation when the ferromagnetic cylinder of the length $L$ and radius $R\gg l_{so}$ is embedded into the bulk of the superconductor (see Fig.~1c). In this case the magnetic field profile is
\begin{equation}\label{Cylinder_H_profile}
H_z(r)=\left\{\begin{array}{l}{H_{F}~~~~~~~~~~~~~~~~~~~~~~~~~~~~{\rm for}~r<R,}\\{ (H_{F}+\Delta H)\frac{K_0\left(r/\lambda\right)}{K_0\left(R/\lambda\right)}~~~~~~{\rm for}~r\geq R.}\end{array}\right.
\end{equation}
Here $K_0$ is the modified Bessel function of the second kind and $H_F$ is the constant corresponding to the minimum of the Gibbs free energy
\begin{equation}\label{Cylinder_Gibbs}
G=\frac{L }{4}\int\limits_0^\infty \left[H_z^2+\lambda^2(\partial_r H_z)^2-2H_zH_{0z}\right]rdr+F_{SO}.
\end{equation}
In Eq.~(\ref{Cylinder_Gibbs}) the SOC-induced surface free energy $F_{SO}$ is determined by Eq.~(\ref{F_SO_general}) where the vector potential $\left. A_\theta\right|_{r=R}=-(H_{F}+\Delta H)RQ/2$ with $Q=2\left(\lambda/R\right)K_1\left(R/\lambda\right)[K_0\left(R/\lambda\right)]^{-1}$ has only angular component. Substituting Eq.~(\ref{Cylinder_H_profile}) into (\ref{Cylinder_Gibbs}) we find:
\begin{equation}\label{Cylinder_Gibbs_res}
G(H_F)=G_2+R^2L\left(H_F^2-2H_F H_{0z}\right)\left(1+Q\right)/8,
\end{equation}
where the value $G_2=-(R^2L/8)(\Delta H^2+2\Delta H H_{0z})Q$ does not depend on $H_F$. Similar to the case of the S/F/S system the minimum of the function $G(H_F)$ formally corresponds to $H_F=H_{0z}$. However, in our doubly-connected geometry the total magnetic flux $\Phi=\int_0^\infty B_z(r)2\pi rdr$ through the ferromagnetic cylinder and the adjacent superconducting layer of the width $\sim \lambda$ is quantized, so that $\Phi=n\Phi_0$, where $\Phi_0=\pi/e$ is the superconducting magnetic flux and $n$ is an integer number. The magnetic induction $B_z(r)$ inside the F-layer is $B_z(r)=H_z(r)+4\pi M_z$, where $M_z$ is the magnetization which is assumed to be uniform. The corresponding values of the magnetic field inside the ferromagnet are
\begin{equation}\label{HF_quantized}
H_F^{(n)}=\left(\Delta H-4\pi M_z+\frac{n\Phi_0}{\pi R^2}\right)\frac{1}{1+Q}-\Delta H.
\end{equation}
The resulting dependence $H_F(H_{0z})$ which realizes the minimum of the Gibbs free energy (\ref{Cylinder_Gibbs_res}) has the form of the staircase (see Fig.~3). The amplitude of the jumps $H_F^{(n+1)}-H_F^{(n)}$ does not depend on the parameters of the SOC and is determined by the radius of the F-cylinder. In the contrast, the position of these jumps is controlled by the SOC: the jump between the ground states with $H_F=H_F^{(n)}$ and $H_F=H_F^{(n+1)}$ occurs at the field $H_{0z}^{(n)}=\left[(\Phi_0/\pi R^2)(n+1/2)-4\pi M_z-Q\Delta H\right]/(1+Q)$. This feature shows the way for the experimental determination of the value $\Delta H$. Note that the step-like behavior of the field $H_F$ reveals for $R\sim\lambda$ while for $R\gg\lambda$ the distance between the steps is negligibly small and the dependence $H_F(H_{0z})$ becomes almost linear.

\begin{figure}[t!]
\includegraphics[width=0.35\textwidth]{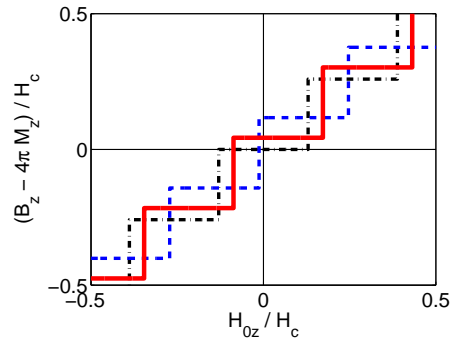}
\caption{(Color online) The dependence of the magnetic field $B_z$ inside the ferromagnetic cylinder as a function of the applied magnetic field $H_{0z}$ (red curve). The dashed blue curve corresponds to the case when there is no spin-orbit coupling in the system, the black dash-dotted curve shows the dependence $B_z(H_{0z})$ for the case $M_z=0$ and $\alpha_{so}=0$. The radius of the F-cylinder is $R=\lambda$, the parameter $\alpha_{so}h_z=0.1\Phi_0/\pi$, $H_c=\Phi_0/\pi\lambda^2$, $M_z=4.5H_c$.}\label{Fig_Staircase}
\end{figure}

To sum up, we predict the emergence of the spontaneous superconducting current at the interfaces between a superconductor and a ferromagnet with strong Rashba spin-orbit coupling. The appearance of such currents results in the induction of the stray magnetic field at the sample edges, local increase of the Cooper pair density near the S/F interface, the changes in the slope of the dependence $H_{c3}(T)$ and the substantial shift of the dependence of the magnetic field inside the ferromagnetic cylinder imbedded into the superconducting sample as a function of the external field.

Experimentally the spontaneous surface current should reveal through the appearance of the magnetic field in the region where the S/F interface comes to the sample edge. In contrast with the stray magnetic field induced by the ferromagnet the described spontaneous field emerges only below the superconducting transition temperature which makes it easily distinguishable in the magnetic measurements. The most appropriate techniques to observe the predicted effects are the scanning SQUID microscope with the single electron spin sensitivity \cite{Vasyukov} or the local probe measurements of the faint magnetic field with the help of the low-energy muon spin spectroscopy \cite{Morenzoni, Bernardo}. Note that the latter method offers the extreme sensitivity to magnetic field of less than $0.1~{\rm G}$ with the depth resolved sensitivity of few nanometers. From Eq.~(\ref{Current_general}) one finds that with the logarithmic accuracy the estimate for the spontaneous magnetic field at the S/F interface gives $\Delta H\sim p_{so}l_{so}H_{c1}(T)$ where $p_{so}=\varepsilon h_z/\gamma$ is the momentum characterizing the shift in the electron energy bands due to SOC, and $H_{c1}(T)=\Phi_0/(4\pi\lambda^2)$ is the lower critical field. Taking $l_{so}\sim 1-10 ~{\rm nm}$ we obtain that the field $\Delta H$ can become of the order of $H_{c1}$ or even exceed it. In the latter case the spontaneous surface current should produce Abrikosov vortices near the S/F interface. Also the emergence of the surface current can be observed in the $H_{c3}$ measurements. At the fixed temperature the difference $\delta H_{c3}$ between the $H_{c3}$ values for the parallel and the antiparallel orientations of the external field and the exchange field in the F-layer is of the order of $\delta H_{c3}/H_{c3}\sim p_{so}l_{so}$. Thus, if $\varepsilon>0$ then for ${\bf h}\uparrow\uparrow {\bf H_0}$ the SOC favors the emergence of the localized superconducting nuclei above the upper critical field $H_{c2}$ while for ${\bf h}\uparrow\downarrow {\bf H_0}$ the formation of such nuclei can even become impossible (if $\varepsilon<0$ the deviation of $H_{c3}$ from $H_{c3}^0$ has the opposite sign). Finally, the spontaneous currents can be observed in the cylinder geometry where for $p_{so}l_{so}\sim 1$ and $R\sim\lambda$ the shift of the steps on the dependence of the magnetic field inside the cylinder on $H_{0z}$ due to SOC becomes of the order of the steps width.

Note that the discovered phenomena should result in a variety of the edge effects such as renormalization of the surface barrier for the Abrikosov vortices, anisotropy of the depairing current in the regime of surface superconductivity, etc. Note also that these effects are not specific for the S/F interfaces with the SOC but should be also relevant for a wide class of interfaces between the superconductors and materials with spin polarization and broken inversion symmetry such as topological insulators or other types of quantum spin-Hall systems.

\begin{acknowledgments}
The authors thank A. Mel'nikov for fruitful discussions and Zh. Devizorova for a valuable comment. This work was partially supported by the
French ANR projects ``SUPERTRONICS'' and ``MASH'', the Russian Foundation for Basic Research (Grant No. 15-02-04116À, calculation of the spontaneous currents in the S/F cylinder) and the Russian Science Foundation (Grant No. 15-12-10020, calculation of the magnetic field profiles in the S/F sandwiches).
\end{acknowledgments}

\break

\section{Calculation of the order parameter profile near the S/F interface}\label{Sec_OrderParameter}

To obtain the spatial distribution of the superconducting order parameter near the S/F interface we rewrite the surface energy in the form $F_{SO}=-(2/3)S\nu|\psi|^3$ with $\nu=48\sqrt{2\pi}\alpha_{so}^2h_z^2e^3\gamma^{3/2}$. Taking the derivative over $\psi^*$ we get the modified Ginzburg-Landau equation:
\begin{equation}\label{GL_modified}
a\psi+\gamma{\bf D}^2\psi+b|\psi|^2\psi-\nu|\psi|\psi\delta(x)=0,
\end{equation}
where $\delta(x)$ is the Delta-function. The main effect of the SOC on the profile $\psi(x)$ is coming from the last term in Eq.~(\ref{GL_modified}) so we may neglect the vector potential arising due to the SOC in the gradient term. Typically, the SOC constant $\nu$ is small so we consider the effect of SOC perturbatively representing the order parameter as the sum $\psi(x)=\psi_0+\psi_1(x)$, where $\psi_0=\sqrt{|a|/b}$ and $\psi_1(x)$ is the correction due to SOC which satisfies the equation $-\gamma\partial^2_x\psi_1+2|a|\psi_1-\nu\psi_0^2\delta(x)=0$. The solution of this equation defines the order parameter profile near the S/F interface. Let us stress that the most favorable conditions for the described growth of the order parameter are realized in the case of ferromagnetic insulators while in the systems with metallic ferromagnets this effect can be masked by the inverse proximity effect which results in the order parameter damping near the S/F interface.

Note that the negligibility of the vector potential in the gradient term in Eq.~(\ref{GL_modified}) takes place only for small values of the spin-orbit coupling constant. Indeed, to estimate the orbital effect of the spontaneous surface current on the distribution of the order parameter we consider the limit $H_{0z}=0$. In this case the vector potential at the interface is determined by the SOC: $A_y\sim \Delta H\lambda$. Demanding that $\gamma e^2A_y^2\psi_0\ll b\psi_0^3$ and taking into account that $\gamma/b\sim\xi^2\psi_0^2$ we finally obtain that the effect of the vector potential is small provided $\Delta H\ll H_{cm}$, where $H_{cm}=\Phi_0/(2\sqrt{2}\pi\xi\lambda)$ is the thermodynamical critical magnetic field of the superconductor. Note that for the type-II superconductors with $\lambda\gg\xi$ this condition can be satisfied even if the magnetic field step $\Delta H$ exceeds the lower critical field $H_{c1}$.

\section{Calculation of the critical field $H_{c3}$}\label{Sec_Supp_Hc3}

In this section we calculate the critical field for the nucleation of the localized states near the surface of the superconductor in the presence of the spin-orbit coupling and the exchange field in the adjacent ferromagnetic layer. Choosing the vector potential gauge $A_y=H_{0z}x$ and following the standard procedure described in Ref.~[A1] we search the solution of the linearized Ginzburg-Landau equation inside the superconductor in the form $\psi(x,y)=f(x){\rm exp}(iky)$ where the function $f$ satisfies the equation
\begin{equation}\label{LGL_Hc3}
\begin{array}{c}{\ds
-\gamma\partial_x^2f+\gamma\left(k-2eH_{0z}x\right)^2f+\alpha(T-T_c)f~~~~~~~~~~~~~}\\{}\\{\ds ~~~~~~~~~~~~~~~~~~~+ 2\varepsilon l_{so} h_z(k-2eH_{0z}x) f \delta(x)=0.}
\end{array}
\end{equation}
Here we neglect the finite thickness of the region with the SOC representing the surface term as the Delta-function. It is convenient to use the dimensionless values $X=x/\xi$, $K=k\xi$ and $g_0=(2eH_{0z}\xi^2)^{-1}$. Then for $T<T_c$ one can represent Eq.~(\ref{LGL_Hc3}) in the form
\begin{equation}\label{LGL_Hc3_Dimless}
-\partial_X^2f+\left(K-X/g_0\right)^2f=f
\end{equation}
with the boundary condition
\begin{equation}\label{LGL_Hc3_BC}
\left.\partial_Xf\right|_{X=0}=sK f(0),
\end{equation}
where $s=2\varepsilon l_{so}h_z/\gamma$.

The value of $H_{c3}$ depends on the mutual orientation of the magnetic field and the exchange field in the ferromagnet. Assuming for definiteness that $H_{z0}>0$ we introduce the new coordinate $t=X/\sqrt{g_0}-K\sqrt{g_0}$ which allows to rewrite Eqs.~(\ref{LGL_Hc3_Dimless})-(\ref{LGL_Hc3_BC}) in the form
\begin{equation}\label{LGL_Hc3_Final}
-\partial_t^2f+t^2f=g_0f,
\end{equation}
\begin{equation}\label{LGL_Hc3_BC_Final}
\left.\frac{\partial_tf}{f}\right|_{t=-\mu}=s\mu,
\end{equation}
where $\mu= K\sqrt{g_0}$.

The solution of Eq.~(\ref{LGL_Hc3_Final}) which does to zero at $t\to\infty$ reads
\begin{equation}\label{LGL_Gen_sol}
f=f_0e^{\frac{t^2}{2}}I_0(-t),
\end{equation}
where $f_0$ is a constant and
\begin{equation}\label{I_def}
I_\beta(u)=\int\limits_0^\infty w^\beta w^{-\frac{1+g_0}{2}}e^{-(w-u)^2}dw.
\end{equation}
Using the relation $\partial_u I_\beta(u)=-2uI_\beta(u)+2I_{\beta+1}(u)$ we rewrite the boundary condition (\ref{LGL_Hc3_BC_Final}) in the form
\begin{equation}\label{BC_I}
2I_1(\mu)-(1-s)\mu I_0(\mu)=0.
\end{equation}
Eq.~(\ref{BC_I}) implicitly defines the function $g_0(\mu)$. The maximal value of the magnetic field $H_{0z}$ which does not destroy the superconductivity corresponds to the minimum of this function. Since Eq.~(\ref{BC_I}) is an identity function its derivative over $\mu$ is zero. Calculating this derivative under the condition $\partial_\mu g_0=0$ we obtain:
\begin{equation}\label{BC_Der}
4I_2(\mu)-2(3-s)\mu I_1(\mu)+(1-s)(2\mu^2-1)I_0(\mu)=0.
\end{equation}
Using the definition of the function $I_\beta(\mu)$ one may check that $4I_2(\mu)=4\mu I_1(\mu)+(1-g_0)I_0(\mu)$. Combining this relation and Eq.~(\ref{BC_I}) we rewrite Eq.~(\ref{BC_Der}) in the form
\begin{equation}\label{BC_Der2}
g_0=s+ (1-s^2)\mu^2.
\end{equation}
Finally, to calculate the critical field we substitute this expression for $g_0$ into Eq.~(\ref{BC_I}) and obtain the integral equation
\begin{equation}\label{Hc3_eq}
\int\limits_0^\infty \left[2w -(1-s)\mu \right]w^{-\frac{1+s+ (1-s^2)\mu^2}{2}}e^{-(w-\mu)^2}dw=0.
\end{equation}
Solving this equation numerically we find that for $s\ll 1$ the value $g_0$ have the form $g_0=\bar{g}_0+ps+O(s^2)$ where $\bar{g}_0=0.5901$ and $p=0.5855$. Going back to the dimensional variables we get:
\begin{equation}\label{g_res}
H_{c3}(T)\approx\frac{\alpha}{2e\gamma}\frac{1}{\bar{g}_0}\left(T_c-T\right)\left[1-\frac{2\varepsilon l_{so} h_z }{\gamma}\frac{p}{\bar{g}_0}\right].
\end{equation}
Thus, when the exchange field inside the ferromagnet and the external field have the same direction ($h_z>0$) the SOC decreases the critical field $H_{c3}$ while in the case when ${\bf H}_0$ and ${\bf h}$ have the opposite directions the SOC results in the increasing of $H_{c3}$.

\bigskip

[A1] D. Saint-James, G. Sarma, and E. J. Thomas, {\it Type II Superconductivity} (Pergamon Press, Oxford, 1969).

\end{document}